\documentclass[11pt]{article}
\usepackage{moriond,epsfig}

\def\be{\begin{equation}}
\def\ee{\end{equation}}
\def\bea{\begin{eqnarray}}
\def\eea{\end{eqnarray}}

\begin{document}
\vspace*{4cm}
\title{Newest results on XYZ states at BESIII}

\author{F. Nerling\footnote{Email: F.Nerling@gsi.de}\\on behalf of the BESIII Collaboration}

\address{Institut f\"ur Kernphysik, Goethe Universit\"at Frankfurt, \\ and  GSI Darmstadt, Germany}

\maketitle\abstracts{
The BESIII experiment at BEPCII at IHEP in Beijing/China has collected the 
world largest data sets in the $\tau$-charm region and it is well suited to 
cover a rich hadron physics programme, including charmonium and open-charm 
spectroscopy, $R$-scan or electromagnetic form factor measurements, and many more.
In particular, the unique data sets between 3.8 and 4.6\,GeV allow for discoveries 
and precision measurements of the interesting charmonium-like (exotic) ``XYZ'' 
states. The recently published precision cross-section measurement for $Y$ states
is discussed as well as the first observation of the $X(3872)$ in radiative decays, 
also the completion of the two observed isospin triplets $Z_{\rm c}(3900)$ and 
$Z_{\rm c}(4020)$ is briefly summarised.  
}

\section{Introduction}
The discovery of the $J/\psi$ in 1974~\cite{discoveryJpsi} triggered investigations in the charmonium region, 
where the $c\bar{c}$ states can successfully described using potential models, for a recent overview see 
e.g.~\cite{reviewMitchel_etal_2016}. Especially below the open-charm threshold, excellent agreement is achieved 
between theory and experiment --- all the predicted states have been observed with the expected properties. The 
situation above the open-charm threshold appears more complicated. There are still many predicted states that 
have not yet been discovered, and, surprisingly, there are quite some unexpected states that have been observed 
since 2003. Well known examples of these so-called charmonium-like (exotic) ``XYZ'' states are the $X(3872)$ 
observed by Belle in 2003~\cite{X3872_belle}, the $Y(4260)$ and the $Y(4360)$, both discovered by 
BaBar~\cite{Y4260_barbar,Y4360_barbar}, or the manifestly exotic charged state $Z_{\rm c}(3900)^\pm$ discovered by 
BESIII in 2013~\cite{Zc3900_besiii}, and shortly after confirmed by Belle~\cite{Zc3900_belle}. 

The BESIII experiment~\cite{besiii} at BEPCII is the latest incarnation of the Beijing Spectrometer (BES) at the Beijing 
Electron-positron Collider (BEPC) that started in 1989, it begun operation in March 2008 after the 
major upgrades were finalised. The multi-purpose detector is well suited to cover a broad hadron physics programme, 
including charmonium and open-charm spectroscopy, $R$ scan measurements or electromagnetic form factors, and many 
others. It has collected the world largest data sets in the $\tau$-charm mass region, and in the ``XYZ'' region above 
3.8\,GeV, BESIII has accumulated unique high-luminosity data sets of about 5\,fb$^{-1}$ in total to explore the 
still-unexplained $XYZ$ states. 

\section{Newest results on XYZ states}
The BESIII experiment is ideally suited to study conventional as well as charmonium-like (exotic) $XYZ$ states.
We have direct access to $Y$ states ($J^{PC}=1^{--}$) in the $e^+e^-$ annihilation, $X$ states are accessed in 
radiative decays, and also, we can study charged as well as neutral $Z_{\rm c}$ states.

\subsection{The $Y$ states --- $e^+e^-$ production of the $J/\psi\pi^+\pi^-$, the $\psi(2S)\pi^+\pi^-$ and the $h_{\rm c}\pi^+\pi^-$ systems }
The $Y(4260)$ and the $Y(4360)$ had been firstly observed using initial state radiation (ISR) decaying to 
$J/\psi\pi^+\pi^-$ and $\psi(2S)\pi^+\pi^-$, respectively, by BABAR~\cite{Y4260_barbar,Y4360_barbar}. 
Based on increased statistics by about a factor of two, the $Y(4260)$ appears to show a somewhat asymmetric 
shape~\cite{Y4260_barbar_update}. In agreement with their previous result based on lower statistics, confirming 
the $Y(4260) \rightarrow J/\psi\pi^+\pi^-$ discovered by BABAR~\cite{Y4260_barbar}, and in contradiction to the 
BABAR result, Belle claims a lower mass peak, the ``$Y(4008)$'', that needs a second coherent resonance (Breit-Wigner) 
shape in addition to the $Y(4260)$ and some incoherent background in order to describe their data~\cite{Zc3900_belle}.  

\begin{figure}[tp!]
\vspace{-0.3cm}
    \begin{center}
     \includegraphics[clip, trim= 45 147 50 200,width=1.0\linewidth]{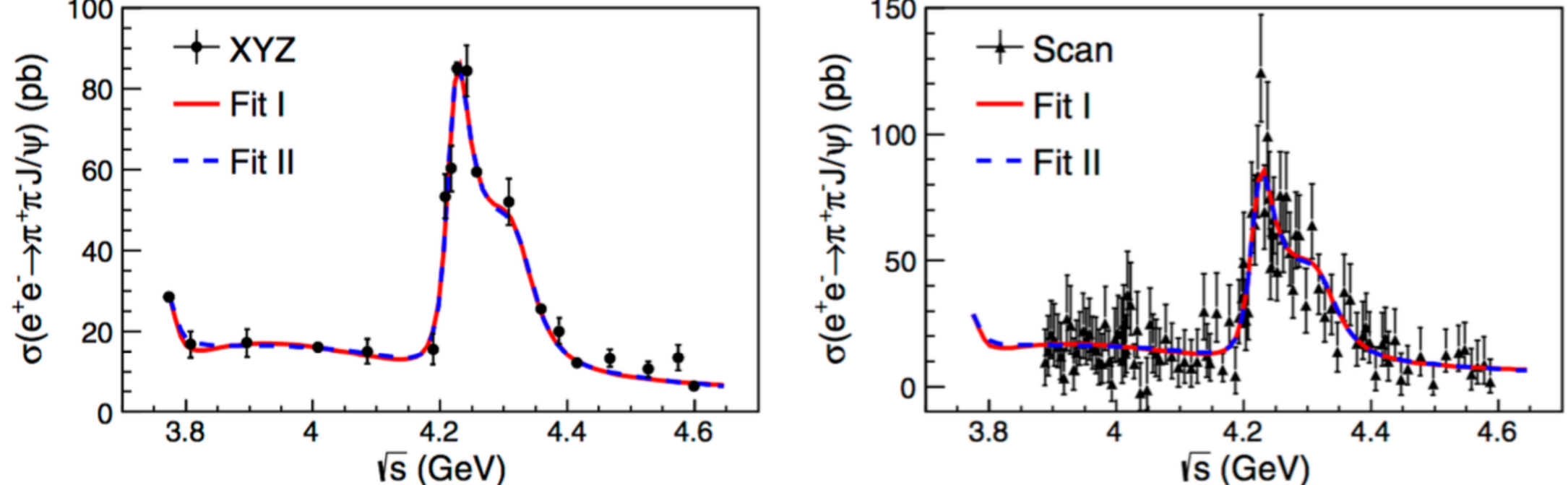}
    \end{center}
\vspace{-0.3cm}
      \caption{Precision cross-section measurement of the $J/\psi\pi^+\pi^-$ production in $e^+e^-$ annihilation as obtained from a 
        simultaneous fit to both, the ``high luminosity'' $XYZ$ data (left) and the ``low luminosity'' scan data 
        (right). 
\vspace{-0.3cm}
}
\label{YstatesJpsipipi_bes3} 
\end{figure}

Based on the ``high luminosity'' ($<$ 40\,pb$^{-1}$ at each energy point $E_{cms}$, total integrated luminosity of 
8.2\,fb$^{-1}$) and the ``low luminosity'' ($\sim$ 7-9\,pb$^{-1}$ at each energy point $E_{cms}$, total integrated 
luminosity of 0.8\,fb$^{-1}$) $XYZ$ data, we performed a precision measurement of the energy dependent cross-section 
$\sigma(e^+e^- \rightarrow J/\psi\pi^ +\pi^-)$ in the energy range of 3.77 $<$ $E_{\rm cms}$ $<$ 4.60\,GeV recently published~\cite{Ycrosssection_besiii}. The result obtained by a simultaneous fit to both data sets is shown in Fig.\,\ref{YstatesJpsipipi_bes3}.
\begin{figure}[bp!]
\vspace{-0.3cm}
    \begin{center}
     \includegraphics[clip, trim= 40 140 60 170,width=1.0\linewidth]{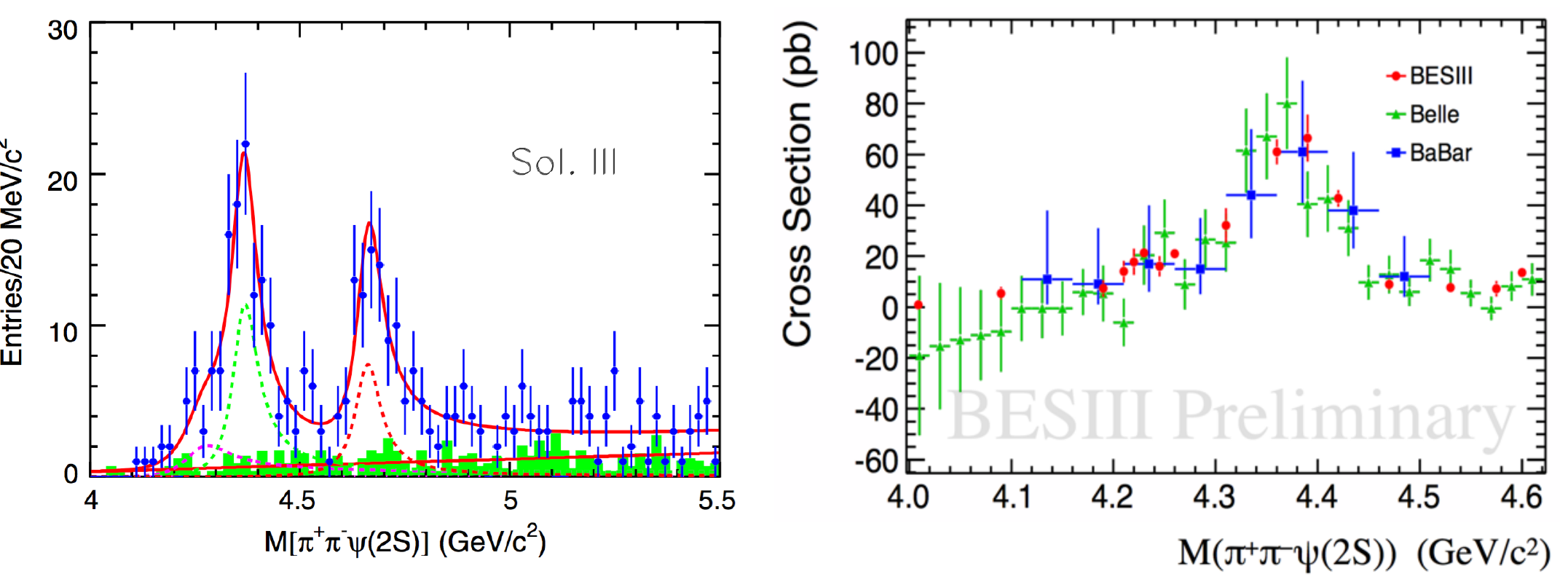}
    \end{center}
    \vspace{-0.3cm}
       \caption{
         {\it Left:} The reconstructed $\psi(2S)\pi^+\pi^-$ invariant mass spectrum from 
         Belle shows clear indications of the $Y(4360)$ and the $Y(4660)$, 
         however, no evidence for the $Y(4260)$ resonance, for details see text.
         {\it Right:} Comparison of the $e^+e^- \rightarrow \psi(2S)\pi^+\pi^-$ cross-section shape 
         measured at BESIII to those provided by BABAR and Belle --- the $Y(4360)$ line shape is 
         found in consistency for the three different experiments.
         }
 \label{Ystates_psi2Spipi_bes3} 
    \vspace{-0.3cm}
\end{figure}
The signal yields are determined using an unbinned maximum-likelihood fit (``high luminosity'' $XYZ$ data) 
and a simple counting method (``low luminosity'' scan data), for the latter the background counts from sidebands 
are subtracted. First of all, the cross-section appears inconsistent with a single peak just for the $Y(4260)$
--- two resonances to describe two peaks is favoured over one by the data at high statistical significance of 
more than $7\,\sigma$. Moreover, two fit models have been used to describe the data. ``Fit I'' comprises three 
resonances, namely $Y(4008)$, $Y(4260)$ and the $Y(4360)$, whereas in ``Fit II'', the first resonance in terms 
of the low mass $Y$ state claimed by Belle has been replaced by an interfering non-resonant exponential shape. 
Since both models deliver identical fit quality, we can not confirm the $Y(4008)$ as it is not needed to 
describe the BESIII data. It should be emphasised that the parameters of the third resonance 
($m=4326.8\pm$10.0\,MeV, $\Gamma=98.2^{+25.4}_{-19.6}$\,MeV) are consistent within errors with the $Y(4360)$ 
observed decaying to $\psi(2S)\pi^+\pi^-$ previously~\cite{Y4360_barbar,psi2Spipi_belle}, while the $Y(4260)$ is 
observed with a significantly smaller width ($\Gamma$=44.1$\pm$3.8\,MeV). Given the much larger statistics by 
BESIII, the $Y(4260)$ and the $Y(4360)$ are resolved here for the first time, and the $Y(4360)$ is first observed 
decaying to $J/\psi\pi^+\pi^-$.      

Interestingly, coming back to the $e^+e^-$ production of the $\psi(2S)\pi^+\pi^-$ system, a result by Belle obtained 
using ISR (Fig.\,\ref{Ystates_psi2Spipi_bes3}, left) gives clear indication of the $Y(4360)$ and the $Y(4660)$ 
decaying to $\psi(2S)\pi^+\pi^-$. However, there is no evidence for the $Y(4260)$ being present in the data. 
When trying to accommodate it coherently, in addition to the other two $Y$ states, the statistical significance 
turns out to be well below 3\,$\sigma$, so that Belle omitted the $Y4260) \rightarrow \psi(2S)\pi^+\pi^-$ from their 
best fit~\cite{psi2Spipi_belle}.  
The preliminary result on the cross-section shape for direct $\psi(2S)\pi^+\pi^-$ production in $e^+e^-$ annihilation 
as observed at BESIII is compared to the ones from BABAR and Belle in Fig.\,\ref{Ystates_psi2Spipi_bes3} (right). 
The preliminary measurement by BESIII confirms the $Y(4360)$ line shape reported previously, they are all three found 
in good agreement. More data for a thorough study of the mass region 4.2--4.3\,GeV is currently taken at IHEP/Beijing.
\begin{figure}[tp!]
\vspace{-0.6cm}
    \begin{center}
     \includegraphics[clip, trim= 45 125 80 200,width=1.0\linewidth]{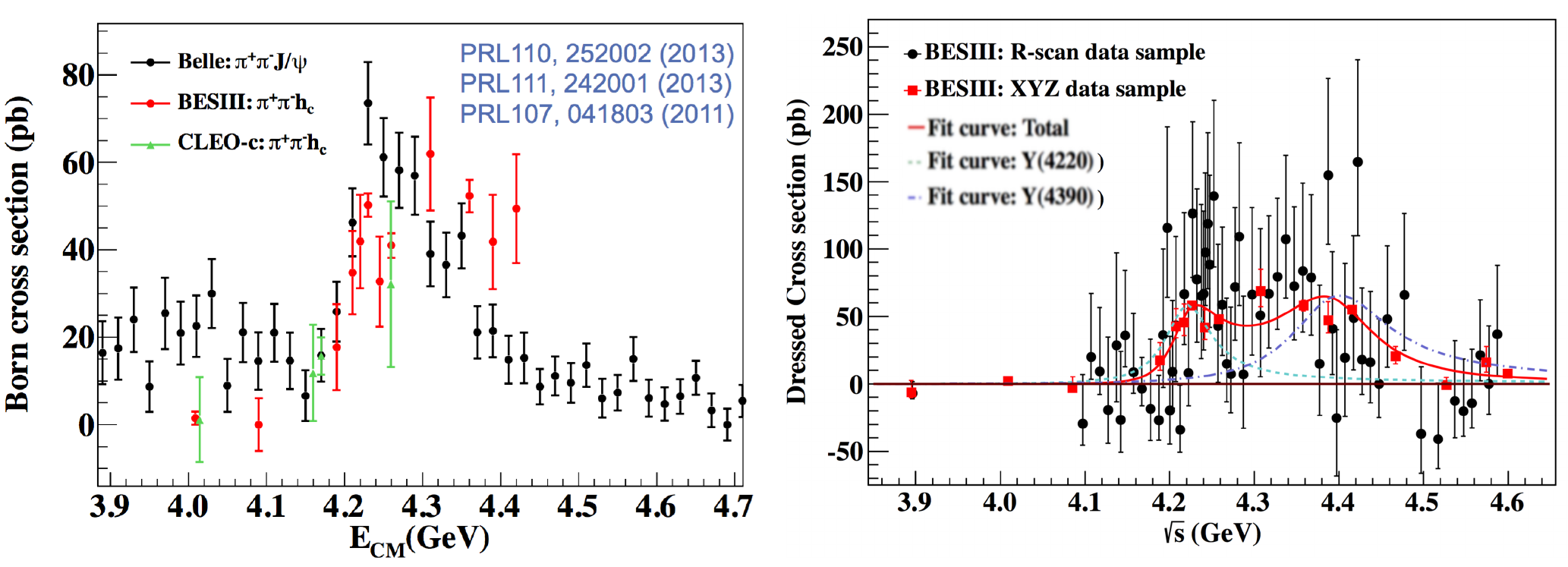}
    \end{center}
      \caption{Cross-section measurement of $h_{\rm c}\pi\pi$ versus $J/\psi\pi\pi$ production 
        in $e^+e^-$ annihilation, based on the ``high luminosity'' $XYZ$ data (left) and based on the ``low luminosity'' 
        scan data (right). 
}
       \label{Ystates_hcpipi_bes3} 
\end{figure}

Finally, the cross-section shapes for $e^+e^-\rightarrow h_{\rm c}\pi^+\pi^-$ production as measured at BESIII and 
CLEO-c are compared to the one for $e^+e^-\rightarrow J/\psi\pi^+\pi^-$ from Belle in 
Fig.\,\ref{Ystates_hcpipi_bes3} (left), and we find the $h_{\rm c}\pi^+\pi^-$ differs in shape from the 
$J/\psi\pi^+\pi^-$ system. 
Based on more statistics (Fig.\,\ref{Ystates_hcpipi_bes3}, right), the new BESIII result~\cite{hcpipi_besiii} shows 
evidence for two resonant structures (at 4.22 and 4.39 GeV/$c^2$) that we call ``$Y(4220)$'' and ``$Y(4390)$'', 
respectively. 
A fit with a coherent sum of two Breit-Wigner functions results in a mass of (4218.4 $\pm$ 4.0 $\pm$ 0.9)\,MeV/$c^2$ 
and a width of (66.0 $\pm$ 0.9 $\pm$ 0.4)\,MeV$c^2$ for the ``$Y(4220)$'', and a mass of 
(4391.6 $\pm$ 6.3 $\pm$ 1.0) MeV/$c^2$ and a width of (139.5 $\pm$ 16.1 $\pm$ 0.6)\,MeV$c^2$ for the ``$Y(4390)$''. 
The statistical significance of ``$Y(4220)$'' and ``$Y(4390)$'' is 10\,$\sigma$ 
over the one resonance assumption.

More work and especially higher statistics data is needed to further sort out these exclusive cross-sections of 
highest interest.
\subsection{The $X$ states --- First observation of $e^+e^- \rightarrow \gamma X(3872)$ }
The $X(3872)$ is the first of the $XYZ$ states observed, it was discovered by Belle in the $J/\psi\pi^+\pi^-$ decay 
mode~\cite{X3872_belle}. At BESIII, we studied $e^+e^-\rightarrow \gamma X(3872) \rightarrow \gamma J/\psi\pi^+\pi^-$
at four centre of mass energies $E_{\rm cms}= (4.009, 4.229, 4.260, 4.360)$\,GeV/$c^2$.
\begin{figure}[tp!]
\vspace{-0.3cm}
    \begin{center}
     \includegraphics[clip, trim= 45 100 80 220,width=1.0\linewidth]{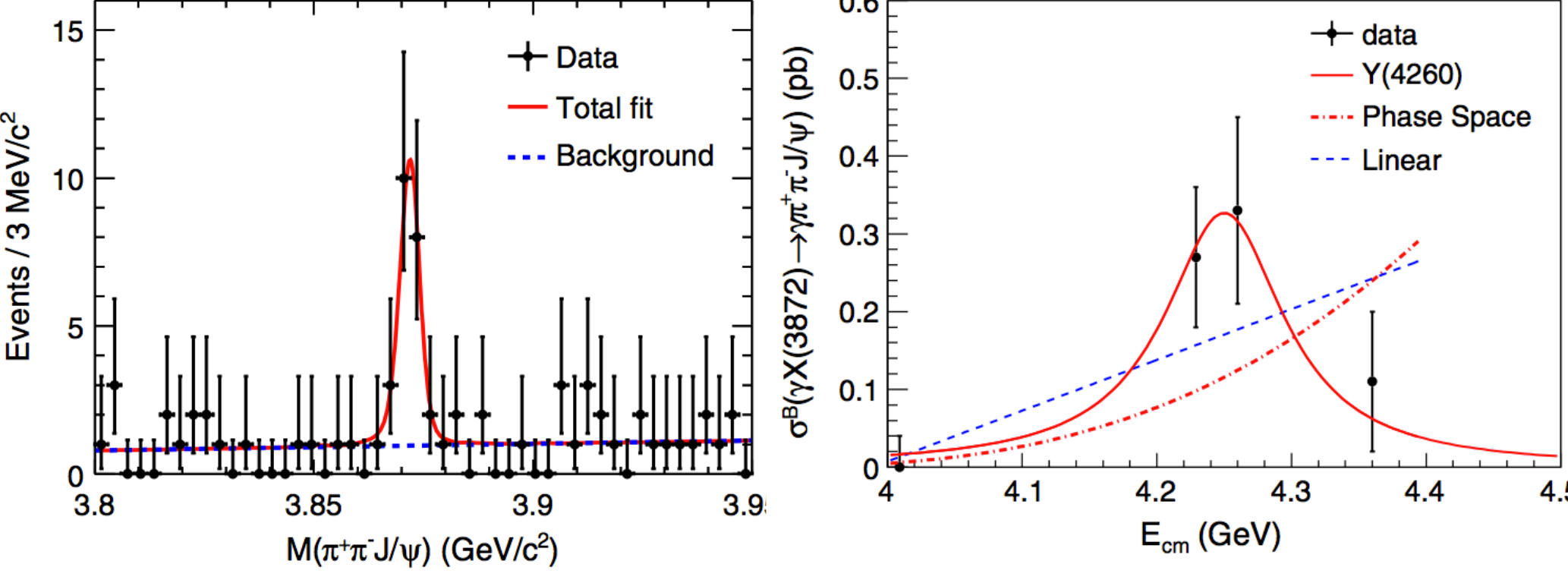}
    \end{center}
      \caption{Observation of the $X(3872)$ produced in radiative decays (left), and the corresponding 
        cross-sections at the four different centre-of mass energies analysed (right). 
\vspace{-0.3cm}
}
       \label{X3872radDecays} 
\end{figure}
The combined reconstructed $J/\psi\pi^+\pi^-$ invariant mass spectrum (Fig.\,\ref{X3872radDecays}, left) shows 
an even though tiny accumulation of events in form of a nice narrow peak on top of a rather low 
background. The resonance parameters obtained from the fit 
($m = (3872.9\pm 0.7 \pm 0.2)$\,MeV/$c^2$, $\Gamma < 2.4$\,MeV/$c^2$) are consistent with those of previous 
observations of the $X(3872)$. This is the first observation of the $X(3872)$ in radiative decays, the 
statistical significance is 6.3\,$\sigma$~\cite{X3872_besiii}. 
The corresponding cross-sections at the four centre-of-mass energies together with different fit attempts 
(Fig.\,\ref{X3872radDecays}, right) already hint to the production via a $Y$ state, however, more data is 
clearly also needed here.

\subsection{The $Z$ states --- Two established isospin triplets $Z_{\rm c}(3900)$ and $Z_{\rm c}(4020)$}
\begin{figure}[tp!]
\vspace{-0.6cm}
    \begin{center}
     \includegraphics[clip, trim= 35 130 30 50,width=1.0\linewidth]{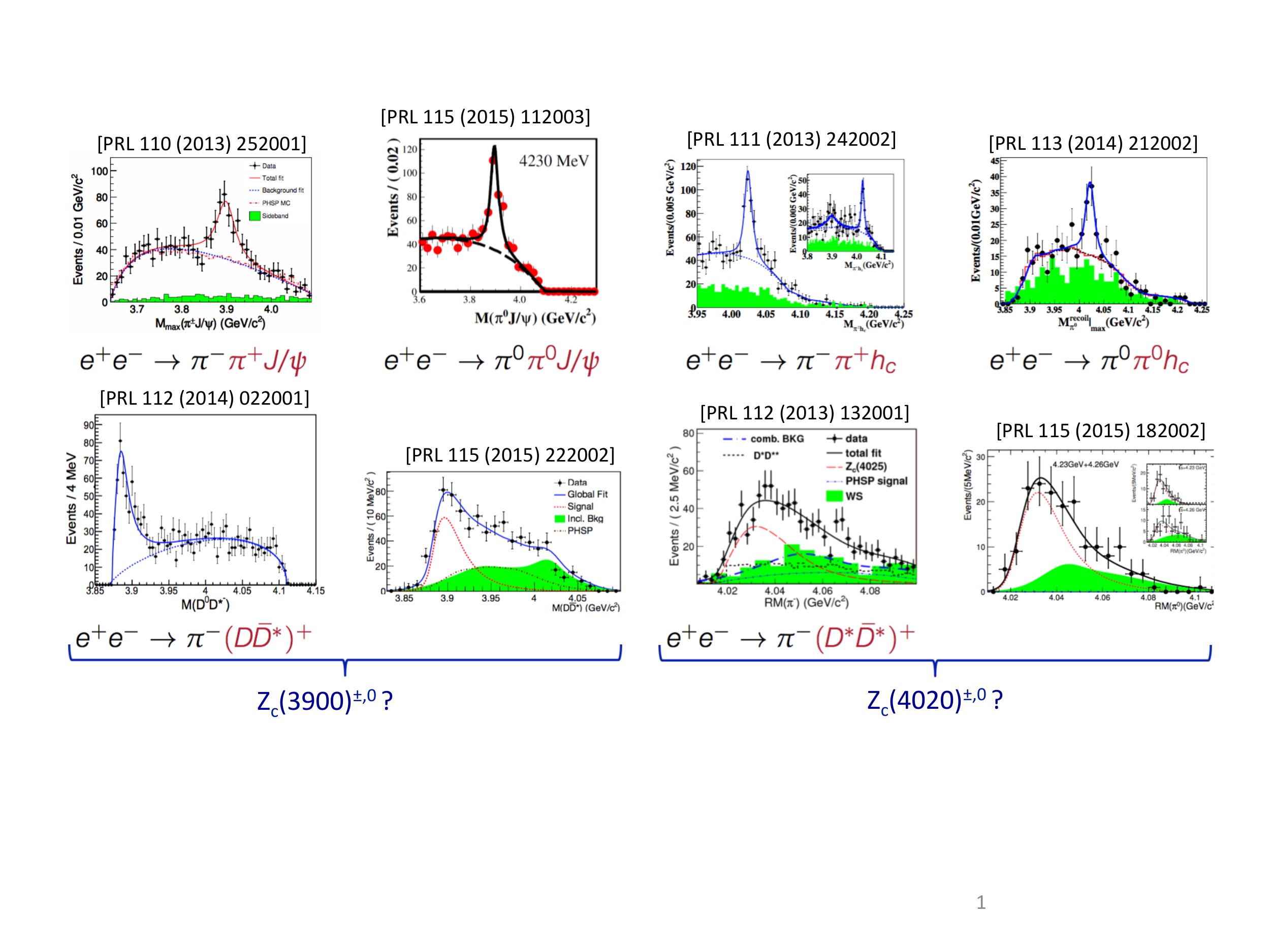}
    \end{center}
      \caption{The two established isospin triplets $Z_{\rm c}(3900)^{\pm,0}$ (left) and $Z_{\rm c}(4020)^{\pm,0}$ (right).
               Shown are the observation plots of the charged (left) and neutral (right) partners, as observed in 
               hidden charm (top) and open charm (bottom) decays.   
}
       \label{twoisospinTriplets_Zc} 
\end{figure}

The $Z_{\rm c}(3900)^\pm$ discovered by BESIII~\cite{Zc3900_besiii}, shortly after confirmed by 
Belle~\cite{Zc3900_belle}, is (due to the charge) a manifestly exotic state, corresponding to a strong hint 
for the first four-quark state being observed. The neutral partner $Z_{\rm c}(3900)^0$ decaying to $J/\psi\pi^0$ 
has been observed also in the BESIII data, at 10.4\,$\sigma$, confirming earlier evidence reported by 
CLEO-c~\cite{cleo-c_ZcNeutral}, and establishing an $Z_{\rm c}(3900)^{\pm,0}$ isospin triplet 
(Fig.\,\ref{twoisospinTriplets_Zc}, left). Furthermore, also a second isospin triplet $Z_{\rm c}(4020)^{\pm,0}$ 
has meanwhile been established in the BESIII data (Fig.\,\ref{twoisospinTriplets_Zc}, right). 
Despite this remarkable progress, the nature of these states is still unclear, especially one question still 
is, whether the different decays the $Z_{\rm c}$ states have been observed in (hidden versus open charm) are decay 
modes of the same state observed. Clearly, also further decay channels via other charmonia (than $J/\psi$ and 
$h_{\rm c}$) need to be investigated, and possible multiplets need to be completed by high spin states, which can 
only be accessed by future experiments like e.g. PANDA at FAIR~\cite{panda}. 
\subsection{Summary and outlook}
The BESIII experiment at BEPCII is successfully operating since 2008. We have collected the world largest data set 
in the $\tau$-charm region as well as unique data sets to study $XYZ$ states. It is ideally suited to explore 
transitions and decays of $Y$ states. We have the first two $Z_{\rm c}$ isospin triplets established, the $X(3872)$ 
observed for the first time in radiative decays, and we have recently published a precision measurement of the 
cross-section in the $Y$ energy range, resolving for the first time structures overseen in previous measurements.
As an outlook, BESIII is continuing collecting data, helping and needed to resolve the $XYZ$ puzzle. 
  
\end{document}